\newif\iffigs\figstrue
\documentclass[12pt]{article}
\usepackage{latexsym,amssymb}
  \iffigs
   \input{epsf}
\else
\fi

\newcommand{\sect}[1]{\setcounter{equation}{0}\section{#1}}

\newcommand{\eq}{\begin{equation}}
\newcommand{\eqa}{\begin{eqnarray}}
\newcommand{\en}{\end{equation}}
\newcommand{\ena}{\end{eqnarray}}
\newcommand{\enn}{\nonumber \end{equation}}

\def\sk{\vskip .4cm}
\def\noi{\noindent}
\def\om{\omega}

\def\al{\alpha}
\def\la{\lambda}

\def\ga{\gamma}
\def\Ga{\Gamma}

\def\Cb{\bar{C}}

\def\rhop{{\rho}^{\prime}}

\def\epsi{\varepsilon}
\def\we{\wedge}

\def\de{\delta}

\def\part{\partial}

\def\R#1#2{ R^{#1}_{~~~#2} }

\def\L#1#2{ \La^{#1}_{~~~#2} }

\def\La{\Lambda}

\def\Cb{{\bf \mbox{\boldmath $C$}}}

\def\c#1#2{ C^{#1}_{~~#2} }
\def\C#1#2{ {\bf \mbox{\boldmath $C$}}^{#1}_{~~#2} }

\def\D{\Delta}

\def\limepsizero{\lim_{\epsi \rightarrow 0}}

\def\n2{{{N+1} \over 2}}

\def\bu{\bullet}
\def\ci{\circ}
\def\square{{\,\lower0.9pt\vbox{\hrule \hbox{\vrule height 0.2 cm
\hskip 0.2 cm \vrule height 0.2 cm}\hrule}\,}}

\def\Q.E.D.{\rightline{$\Box$}}

\def\sumong{\sum_{g \in G}}

\def\sumongnote{\sum_{g \not= e}}

\def\sumonh{\sum_{h \in G}}
\def\sumonhp{\sum_{h \in G'}}

\def\Lcal{{\cal L}}
\def\Rcal{{\cal R}}

\def\omc#1#2{\om^{#1}_{~~#2}}
\def\W#1#2{W^{#1}_{~~~#2} }

\def\Gc#1#2{\Ga^{#1}_{~~#2}}
\def\a#1#2{a^{#1}_{~#2}}
\def\ainv#1#2{(a^{-1})^{#1}_{~#2}}

\def\unop{1  \! \mbox{I}}

\def\ph{\hat p}
\def\qh{\hat q}

\def\dleft{\stackrel{\leftarrow}{\partial}}
\def\dright{\stackrel{\rightarrow}{\partial}}
\def\D{\triangle}

\def\Amu{A_{\mu}}
\def\Anu{A_{\nu}}
\def\Abu{A_{\bu}}
\def\Atimu{\Ati_{\mu}}

\def\Atibu{\Ati_{\bu}}
\def\Fmunu{F_{\mu\nu}}
\def\Fmubu{F_{\mu\bu}}

\def\Fbubu{F_{\bu\bu}}
\def\fti{{\tilde f}}
\def\Ati{{\tilde A}}
\def\Uti{{\tilde U}}
\def\Gti{{\tilde G}}
\def\Jti{{\tilde J}}
\def\psiti{{\tilde \psi}}
\def\psibar{{\bar \psi}}
\def\chibar{{\bar \chi}}

\begin{document}
\begin{titlepage}
\vskip -1cm \rightline{DFTT-20/2000}
\rightline{May 2000} \vskip 1em
\begin{center}
{\large\bf Noncommutative geometry and physics: a review of
selected recent results }
\\[2em]
Leonardo Castellani
\\[.7em] {\sl Dipartimento di
Scienze e Tecnologie Avanzate,
 East Piedmont University, Italy;} \\  {\sl
Dipartimento di Fisica Teorica and Istituto Nazionale di Fisica
Nucleare\\ Via P. Giuria 1, 10125 Torino, Italy.} \\
{\small castellani@to.infn.it}\\[2em]
\end{center}
\vskip 4 cm
\begin{abstract}
\sk This review is based on two lectures given at the 2000 TMR
school in Torino$^*$. We discuss two main themes: i) Moyal-type
deformations of gauge theories, as emerging from M-theory and open
string theories, and ii) the noncommutative geometry of finite
groups, with the explicit example of $Z_2$, and its application to
Kaluza-Klein gauge theories on discrete internal spaces.

\end{abstract}

\vskip 5cm
 \hrule
  {\centerline{\small $^*$ TMR school on contemporary String Theory and Brane
 Physics}}
 {\centerline {\small January 26 - February 2, 2000, Torino, Italy.}}
  \vskip 1cm
   \noi{\centerline {\footnotesize
  Supported in part by   EEC  under TMR contract
 ERBFMRX-CT96-0045}}

\end{titlepage}
\newpage
\setcounter{page}{1}

\sect{Introduction}

Noncommutativity of coordinates is not a surprising occurrence in
physics, quantum phase space being the first example that comes to
mind. In fact some early considerations on its ``quantized"
differential geometry can be found in \cite{dirac}. This
particular operator algebra has inspired the idea of {\sl
spacetime} coordinates as noncommuting operators.
 The idea has been explored since quite some time \cite{snyder} in various
 directions, one main motivation being that the relation:
 \eq
 [x^{\mu},x^{\nu}]=i~\theta^{\mu\nu} \label{xxcomm}
 \en
 embodies an uncertainty principle that smears the spacetime picture
 at distances shorter than $\sqrt{\theta}$, and therefore a natural
 cutoff when using a quantum field theory to describe natural
 phenomena.
 Since ``measuring" spacetime geometry under
 distances smaller than the Planck length $L_P$ is not
 accessible even to Gedanken experiments
(at this scale the curvature radius of spacetime becomes of the
order of a probe particle wavelength), relation (\ref{xxcomm})
seems to make good physical sense when  $\sqrt{\theta} \approx
L_P$. Thus a quantum theory of gravity containing or predicting
relation (\ref{xxcomm}) would have a good chance to be
intrinsically regulated.
 \sk
String theories have been pointing towards a noncommuting scenario
already in the 80's \cite{wittensft}. More recently Yang-Mills
theories on noncommutative spaces have emerged in the context of
$M$-theory compactified on a torus in the presence of constant
background three-form field \cite{CDS}, or as low-energy limit of
open strings in a backround $B$-field
\cite{douglashull}-\cite{seibergwitten},
describing the fluctuations of the $D$-brane world volume . As
observed for example in \cite{schomerus}, noncommutativity in open
string theories is to be expected at some level, since open string
vertex operators are inserted along a one-dimensional line, i.e.
the boundary of the world sheet: the points of insertion are
canonically ordered, so that the product of two such operators
depends on their order of insertion. For a comprehensive account
on noncommutativity in string theory and M-theory we refer to D.
Bigatti's lectures \cite{bigatti}, and to earlier reviews (for ex.
\cite{douglas}).

 The first part of this review  concerns
 a short description of noncommutative Yang-Mills
theories, with emphasis on the algebraic structure, that is on the
(noncommutative) Moyal product, and with some remarks on the relations
between deformed products and quantization rules. Recent results on
perturbative aspects of noncommutative scalar field theories are recalled.

The second part
is devoted to the differential geometry of finite
groups. The general theory is illustrated in the case of $Z_2$. As
a physical application, we construct a gauge theory on $M_4 \times
Z_2$, obtaining via a Kaluza-Klein mechanism a Higgs field (with
the correct spontaneous symmetry-breaking potential and Yukawa
couplings) in  $d=4$ Minkowski spacetime $M_4$.
 \sk
Noncommutative geometry (NCG) has a vast literature that we do not
even attempt to cite. Reviews can be found
in \cite{connesbook,landi,varillyreview,madorebook}. We just mention some of its uses in physics
not discussed in these two lectures: Connes' program of
reconstructing the standard model from the NCG of suitable
operator algebras \cite{connes}; quantum groups, i.e. continuous
deformations of Lie groups, and their NCG
applied to gauge and gravity theories (see for ex.
 \cite{AC,LCqgauge,LCqgravity}), deformed quantum mechanics
and solid state physics.
 \sk

 To find the geometry corresponding to a given algebraic structure
 is a fascinating and usually difficult task, whereas the
 inverse route is often much easier. A constructive starting point
 for NCG is to reformulate as much as possible the geometry of
 a manifold in terms of an algebra of functions defined on it
 \footnote{For example tangent vectors on a manifold $V$
 can be seen as derivations on the functions on $V$, etc. }, and
 then to generalize the corresponding results of differential
 geometry to the case of a noncommutative algebra of functions.
 The main notion which is lost in this generalization is that of a
 {\sl point} (``noncommutative geometry is pointless geometry").

\sect{From sets of points to algebras of functions: $C^*$
algebras}

 The primordial arena for geometry and topology are sets $V$ of points
 with some particular structure. Such a set we call ``space". In
 many cases this set is completely characterized by an algebra of
 functions on it, so that all the information about $V$ can be
 retrieved from the functions alone.

 Let us start with an elementary example: a finite dimensional
 vector space $V$. The functionals
 \eq
 f: V \rightarrow \mathbb{R}~ or~ \mathbb{C}
 \en
constitute the dual vector space $V^*$ isomorphic to $V$, a basis
in $V^*$ being given by the functionals $x^i$, dual to the basis
vectors $v_j$ of $V$: $x^i (v_j)=\de^i_j$. The study of $V^*$ is
completely equivalent to the study of $V$.
 \sk
More generally consider a set $V$ of points, and the algebra of
complex valued functions on $V$, $A = Fun(V)$. This algebra is
clearly associative and commutative, with the usual pointwise
product and sum: $(f \cdot g)(v)=f(v) g(v),~(f +g)(v)=f(v)+ g(v),~
(\la f)(v)=\la f(v), \la \in \mathbb{C}$. The unit $I$ of the
algebra is given by the function $I(v)=1, \forall v \in V$. As a
simple example suppose again that $V$ has a finite number of
elements. Then $A$ is of finite dimension as a vector space, and
any $f \in A$ is expressible as
 \eq
 f=f_i x^i, ~~x^i (v_j) = \de^i_j
 \en
 where now $v_j$ are the elements of $V$.
 Note the multiplication rule:
 \eq
 x^i x^j = \de^{ij} x^i
 \en
 A norm can be defined in $A$ : $\|f\| \equiv max_{v \in V}
 |f(v)|$. Let $f^*$ be the complex conjugate of $f$, then
 \eq
 \|f~f^* \|=\|f\|^2 \label{normC}
 \en
 A normed algebra with an involution $f \rightarrow f^*$
 satisfying (\ref{normC}) is called a $C^*$ algebra. Thus
 $A=Fun(V)$ is a (commutative) $C^*$ algebra.
 \sk
 Conversely any $n$-dimensional commutative $C^*$ algebra
 can be considered
 as algebra of functions on a set of $n$ points. Note that
 commutativity is essential to interpret it as an algebra of
 functions on a set of points.
 \sk
 The finite dimensional example extends to infinite sets if they
 have a topology. In fact if $V$ is a compact space,  then the
 algebra $C^{\ci}(V)$ of continuous functions on $V$ is a $C^*$
 algebra. Conversely any $C^*$ algebra $A$ with a unit element is
 isomorphic to the algebra of continuous complex functions on some
 compact space $V$. This space is just the space of homomorphisms
 $\chi$ from $A$ to $C$ such that $\chi (I) = 1$. The points of
 $V$ are then in 1-1 correspondence with  irreducible
 representations of $A$. This is essentially the commutative
 Gel'fand-Naimark theorem.
 \sk
 Replacing now the commutative $A$ with a noncommutative $A$, the
 ``space" may be hard to find: in most cases these algebras have
 non nontrivial homomorphisms into $\mathbb{C}$, so that the
 reconstruction of a space fails. But the existence of such a
 space may not be necessary, if one has transferred all the
 relevant information for a physical theory into the algebra $A$.
\sk
 There are various ways to generalize to the noncommuting case.
Continuous deformations of commutative $A$ into noncommutative $A$
include quantum groups (and quantum coset spaces) and deformations
of Poisson structures, of which the noncommutative torus is a
simple example. In these cases there is a set of continuous
parameters that control the noncommutativity, and one recovers the
commuting case (the ``classical limit") for some specified values
of these parameters.

On the other hand there are noncommutative algebras that are not
connected to a commutative limit, as in the case of matrices with
entries in $Fun(V)$. An example that we will work out in some
detail in Section 4 is the differential geometry of finite groups:
in this case $Fun(V)$ is commutative, but the differentials do not
simply anticommute between themselves and do not commute with
functions: hence a noncommutative {\sl differential} geometry.

\sect{Deformation quantization}

Consider the algebra of smooth functions on phase space.
Deformation quantization essentially consists in deforming the usual
commutative product between functions into an  associative
 noncommutative product, the ``star" product:
\eq
A * B = AB + i {\hbar \over 2} \{ A,B\}_{PB} + 0(\hbar^2)
\label{starproduct}
\en
where $\hbar$ is a parameter ($\hbar \rightarrow 0$ corresponds
to the commutative limit), and $\{ A,B\}_{PB}$ is the Poisson bracket
of the two phase space functions $A(q,p),B(q,p)$. Imposing associativity
of the star product determines the higher $0(\hbar^2)$ terms
up to equivalences that we discuss in next paragraph.

More generally, on a given manifold $X$ with a Poisson structure
there is essentially one star product, modulo gauge equivalences
that amount to linear redefinitions of the functions:
\eq
A \rightarrow D(\hbar)A \equiv A + \hbar D_1 (A) + \hbar^2 D_2 (A)
+...,
\label{stargauge}
\en
$D_i:Fun(X) \rightarrow Fun(X)$ being differential operators.
This result was proved, in the sense of formal series expansions,
in ref.s \cite{kontsevich,fedosov}. That the linear automorphisms
(\ref{stargauge}) are gauge transformations with respect to the
star product can be understood as follows: consider the
deformation of the ordinary product $AB$ due to $D(\hbar)$: \eq
A*B = D(\hbar)^{-1} (D(\hbar)(A) D(\hbar)(B))
\en
This product is still commutative, and not essentially different
from the ordinary one. Two $*$
products related by $D(\hbar)$ may therefore be considered equivalent.
Thus deformation quantization yields a noncommutative algebra
of functions for each Poisson structure on the manifold $X$.
Poisson structures $\{~,~\}$ can be parametrized by an antisymmetric
 tensor $\theta^{ij}(x)$
such that $\{A,B\} \equiv \theta^{ij}(x) (\part_i A) (\part_jB)$,
satisfying differential identities corresponding to the Jacobi
identities of the Poisson bracket. The simplest Poisson structure
is of course the Poisson bracket of ordinary (flat) phase-space,
whose noncommutative algebra we consider in the following.
 \sk
  Historically the deformations (\ref{starproduct})
 arose in studying the noncommutative
structure of quantum mechanics, and this explains the word
``quantization" and the appearance of the symbol $\hbar$ as deformation
parameter. Consider for example the Weyl
quantization rule $W$ (a linear map from the
classical phase-space functions to the quantum operators)
of the basic phase space monomial:
\eq
q^m p^n \rightarrow W(q^m p^n)= {1 \over 2^n} \sum_{k=0}^{n}
\left( \begin{array}{c}
  n \\
  k
\end{array} \right) \ph^{n-k} \qh^m \ph^k
\en
where $\qh,\ph$ are the quantum phase space operators. This rule
amounts to sum on the permutations of all $\ph$ and $\qh$ considered
as different objects, thus producing an hermitian operator. For example
\eq
W(qp^2) = {1\over 4}(\ph^2 \qh + 2 \ph\qh\ph + \qh \ph^2)
\en
Note that this rule can be efficiently restated as
\eq
W(q^m p^n)= \left[ \exp [-{1\over 2} i\hbar ({\part^2\over \part q \part p})]
                   q^m p^n  \right]_{q \rightarrow \qh,p \rightarrow
                   \ph}
\en
where the substitution $q \rightarrow \qh,p \rightarrow \ph$
occurs on each monomial $q^r p^s$ with $q$'s ordered to the left.
This formula may be checked to hold on the basic monomial, and
extends therefore to any phase-space function $A(q,p)$ expressible
as a power series:
 \eq
  W(A(q,p))=:\exp [-{1\over 2} i\hbar
({\part^2\over \part q \part p})]A(q,p):
 \en
$:~~:$ indicating normal ordering ($q$ preceding $p$) and
substitution $q \rightarrow \qh,p \rightarrow \ph$. The  map $W$
is invertible, i.e. there is a 1-1 correspondence between quantum
operators and functions on phase-space. This is essentially the
core of Moyal formalism \cite{moyal,uhlhorn}, enabling the study
of quantum systems within the classical arena of phase-space via
the inverse map $W^{-1}$.
 \sk
  Consider the product of two
quantum operators $W(A)$, $W(B)$: the classical image $W^{-1}$ of
their product is what is called the Moyal product $A * B$, and is
given by \footnote{In fact the product was introduced by H.
Groenewold \cite{groenewold} (and even earlier, less explicitly,
by J. von Neumann \cite{vonN}). We thank C. Zachos for bringing
this to our attention.}
 \eq
  W^{-1} ( W(A) W(B)) \equiv A*B = A(q,p) \exp [ i
{\hbar \over 2} \D] B(q,p) \label{moyalproduct}
\en
where $\D$ is the bidifferential operator defining the Poisson
bracket:
\eq
A \D B \equiv \{A,B \}_{PB}
\en
i.e. $\D = ({\dleft \over \part q }{\dright \over \part p}
 - {\dleft \over \part p }{\dright \over \part q}) $.
Clearly the Moyal product inherits the properties of the operator
product, i.e. it is associative and noncommutative (unless
the operators $W(A), W(B)$ happen to commute), and gives an
explicit instance of the star product (\ref{starproduct}).
\sk
The Moyal bracket $\{A,B \}_M$
is given by the commutator:
\eq
\{A,B\}_M \equiv A*B-B*A=2i A \sin [ {\hbar \over 2} \D] B
\en
and obviously has all the properties of a Lie bracket: it is
bilinear, antisymmetric and satisfies Jacobi identities. The
Moyal bracket is the image in classical phase-space
of the commutator between quantum operators:
\eq
\{A,B\}_M= W^{-1} ( [W(A), W(B)])
\en
cf. eq. (\ref{moyalproduct}).

Of course the Weyl map is not the only possible quantization rule.
A classification of quantization rules and the construction of the
corresponding noncommutative $*$ products and brackets can be found
in \cite{cas78}. In fact different quantization rules correspond
to $*$ products connected by the gauge transformations
(\ref{stargauge}).
 \sk
  Similarly we can introduce
noncommutativity
 in ordinary $\mathbb{R}^d$ spacetime via a new product on the $C^*$ algebra
 of $C^{\infty}$ complex functions:
 \eq
 A * B (x) \equiv A(x) \exp [{i\over 2} \dleft_{\mu} \theta^{\mu\nu}
 \dright_{\nu} ] B(x) \label{starx}
 \en
where $\theta^{\mu\nu}$ is constant, real and antisymmetric. Then the
commutator of the coordinates $x^{\mu}$ computed with the
star product yields precisely relation (\ref{xxcomm}). By a
change of coordinates $\theta$ can be reduced to the symplectic form:
\eq
\theta=
\left( \begin{array}{rrr}
  0 & 1 &  \\
  -1 & 0 &  \\
   &  & \ddots
\end{array} \right)
\en
Thus if $\theta$ has rank $r$ the relations (\ref{xxcomm}) describe
a spacetime with ${r\over 2}$ pairs of noncommuting coordinates
(with the same algebraic structure as an $r$-dimensional phase-space)
 and $d-r$ coordinates that commute with all the others. In the
 $r$-dimensional subspace the star
 product coincides with the Moyal product discussed previously.
 \sk
 A noncommutative torus is obtained by considering periodic
 coordinates $ 0 \leq x^{\mu} < 2 \pi$. In the periodic case
 it is convenient to redefine the star product (\ref{starx})
 as $A * B (x) \equiv A(x) \exp [ \pi i \dleft_{\mu} \theta^{\mu\nu}
 \dright_{\nu} ] B(x)$ (which amounts to multiply $\theta$ by
 $2\pi$), and to change variables:
 \eq
 U_{\mu} \equiv e^{i x^{\mu}}
 \en
 The product between these new variables becomes:
 \eq
 U_{\mu} * U_{\nu} = e^{ \pi i \theta^{\mu\nu}} e^{i(x^\mu + x^\nu)} =
 e^{2\pi i \theta^{\mu\nu}} U_{\nu} * U_{\mu}
 \en
 Notice that two noncommutative tori related by $\theta^{\mu\nu}
 \rightarrow \theta^{\mu\nu} + \Lambda^{\mu\nu}$, where
 $\Lambda^{\mu\nu}$ is antisymmetric with integer entries,
 are equivalent.
 \sk

  Quantum field theories on noncommutative spacetime (for a very
partial list of ref.s see \cite{filk}-\cite{MST}) are then
obtained by considering their ordinary action and replacing the
usual product between fields with the $*$ product. Indeed the
algebra of functions on noncommutative $R^d$ can be viewed as the
algebra of ordinary functions on the usual $R^d$ with a
deformed $*$ product. Thus we transfer the noncommutativity of
spacetime to the noncommutativity of the product between
functions, and then apply the usual perturbation theory.
Because of the nonpolynomial character of the star product the
resulting field theory is nonlocal. This kind of theories is under
active study. We'll mention here only a few results.

Noncommutative scalar theories at the perturbative level have been
investigated for example in \cite{minwalla}. The quadratic part of
the action is the same as in the noncommutative theory, since
$\int d^dx \phi
* \phi=\int d^dx \phi \phi$ and likewise for the kinetic term
(assuming suitable boundary conditions on $\phi$ that allow to
drop total derivatives). Therefore propagators are the same as in
the commutative case. The interactions however are modified: in
momentum space an interaction vertex $\phi^n$ gives rise to an
additional phase factor:
 \eq V(k_1,k_2,...,k_n)=e^{-{i\over 2}
\sum_{i < j} k_i \times k_j}
 \en
where $k_i$ is the momentum flowing into the vertex through the $i$th
$\phi$ and $k_i \times k_j \equiv (k_i)_{\mu} \theta^{\mu\nu}
(k_j)_{\nu}$. This is the only modification to the Feynman rules
 and its consequences have been investigated in \cite{minwalla},
 finding that $\theta$ dependence factorizes in planar graphs
 (i.e. the phase factor associated with the planar diagram does not
  contain internal momenta),
 yielding then essentially the same behaviour as in the noncommutative theory.
Interesting differences arise in the non-planar diagrams: the
one-loop diagrams turn out to be finite at generic values of the
external momenta,  a consequence of the rapid oscillations of phase
factors of the type $e^{ip \times k}$ where $p$ is an external
momentum and $k$ is the loop momentum. These factors disappear
when $p_{\mu}
 \theta^{\mu\nu} \rightarrow 0$, and the nonplanar graphs become singular in this limit.
 This can be interpreted as a mixing between  UV and IR divergences:
 turning on $\theta$ replaces the UV divergence with a $p \rightarrow
 0$ IR divergence. Moreover, the commutative limit $\theta
 \rightarrow 0$ is not smooth \cite{MST}.

 The quantum analysis is more complicated for noncommutative Yang-Mills
 theories. Classically the noncommutative $U(N)$ YM action is :
 \eq
 S= {1\over 4 g^2} \int Tr(F_{\mu\nu} * F_{\mu\nu})
 \en
 where
 \eqa
& & F_{\mu\nu} = \part_{\mu} \Anu - \part_{\nu} \Amu -i (\Amu *\Anu-
 \Anu * \Amu)\\
 & & A = A^a t^a, ~~~~ Tr(t^at^b)=\de^{ab}
 \ena
As already noticed in \cite{FFZ}, the noncommutative gauge
transformations: \eqa & & \de_{\epsi} \Amu=\part_{\mu} \epsi
-i(\Amu * \epsi - \epsi * \Amu)\\ & &\de_{\epsi} \Fmunu= -i(\Fmunu
* \epsi - \epsi * \Fmunu) \ena leave the action invariant. The
perturbative quantum theory is the object of current research.
 \sk
  The extension of the AdS/CFT
correspondence to backgrouds with constant $B$ field can shed some
light on the nonperturbative regime of noncommutative field
theories, see for ex. ref. \cite{alishahiha}.
 \sk
  Deformation quantization
has been applied to numerous other physical systems,
besides scalar and gauge theories and their supersymmetric
versions \cite{morzum,FL}. We mention for example gravity \cite{antonsen} and the
bosonic string action \cite{garciacompean}.

\sect{Dynamics on finite groups from their noncommutative
geometry}

In this Section we present a systematic method for constructing
field theories on finite groups. This method is based on the
(noncommutative) differential geometry of finite groups, studied
in ref.s \cite{DMGcalculus,FI1,castDCFG,castGFG}. The general
theory is applied to the simplest possible finite group, i.e.
$Z_2$. The example of the simplest {\sl nonabelian} finite group $S_3$
can be found in \cite{castGFG,castDCFG}, and in \cite{castGFG} a
gravity-like theory on $S_3$ is discussed. Here we will
 use the NCG on $Z_2$ to formulate a $U(N)$ gauge theory coupled
 to Dirac fermions on
 $M_4 \times Z_2$,
yielding in $M_4$ (Minkowski spacetime)
 a Yang-Mills theory coupled to Dirac matter
plus a Higgs sector with symmetry-breaking potential
and Yukawa couplings to the fermions.
\sk

Differential calculi can be constructed on spaces
that are more general than differentiable manifolds.
Indeed the general algebraic construction of differential calculus in terms of Hopf
structures \cite{Wor} allows to extend the usual differential geometric
quantities (connection, curvature, metric, vielbein etc.) to a
variety of interesting spaces that include quantum groups,
noncommutative spacetimes (i.e. quantum cosets), and discrete
spaces.
 \sk
 In this lecture  we concentrate on the differential
 geometry of finite group ``manifolds". As
discussed in \cite{DMGcalculus,castGFG,castDCFG}, these spaces can
be visualized as collections of points,
corresponding to the finite group elements, and connected by
oriented links according to the particular differential calculus
we build on them. Although functions $f \in Fun(G)$ on finite
groups $G$ commute, the calculi that are constructed on $Fun(G)$
by algebraic means are in general noncommutative, in the sense
that differentials do not commute with functions, and the exterior
product does not coincide with the usual antisymmetrization of the
tensor product.
 \sk
 Among the physical motivations for finding
differential calculi on finite groups we mention
the possibility of using finite group spaces
as internal spaces for Kaluza-Klein compactifications of
Yang-Mills,
supergravity or superstring theories (
for example Connes'
reconstruction of the standard model in terms of noncommutative
geometry \cite{connes}  can be recovered as Kaluza-Klein
compactification of Yang-Mills theory on an appropriate discrete
internal space). Differential calculi on discrete
spaces can be of use in the study of integrable models, see for ex. ref.
\cite{DMintegrable}. Finally gauge
and gravity theories on finite group spaces may be used as lattice
approximations. For example the action for pure Yang-Mills $\int F
\we {}^* F$ considered on the finite group space $Z^N \times Z^N
\times Z^N \times Z^N$, yields the usual Wilson action of lattice
gauge theories, and $N \rightarrow \infty$ gives the continuum
limit \cite{DMgauge}. New lattice theories can be found by choosing
different finite groups.

While the construction of the differential calculus on finite
groups in ref.s  \cite{DMGcalculus,FI1,castGFG,castDCFG} uses the
Hopf algebraic approach of Woronowicz \cite{Wor}, here this
calculus will be presented without recourse to Hopf algebra
techniques. Most of the content of next Section can be found in
\cite{castGFG,castDCFG}, and is included here for self-consistency.

\subsection{Differential calculus on finite groups}

    Let $G$ be a finite group of order $n$ with
generic element $g$ and unit $e$. Consider $Fun(G)$, the set of
complex functions on $G$. An element $f$ of $Fun(G)$ is specified
by its values $f_g \equiv f(g)$ on the group elements $g$, and can
be written as
 \eq
  f=\sum_{g \in G} f_g x^g,~~~f_g \in \Cb \label{fonG}
 \en
where the functions $x^g$ are defined by
\eq
x^g(g') = \de^g_{g'}
\en
Thus $Fun(G)$ is a n-dimensional vector space, and the $n$ functions
$x^g$ provide a basis. $Fun(G)$ is also a commutative algebra,
with the usual pointwise sum and product, and unit $I$ defined
by $I(g)=1, \forall g \in G$. In particular:
\eq
x^g x^{g'}=\de_{g,g'} x^g,~~~\sumong x^g = I \label{mul}
\en
 Consider now the
left multiplication by $g_1$:
 \eq L_{g_1}g_2=g_1g_2,~~~\forall
g_1,g_2 \in G
\en
This induces the left action (pullback) $\Lcal_{g_1}$ on
$Fun(G)$:
\eq
\Lcal_{g_1} f(g_2) \equiv f(g_1g_2)|_{g_2},~~~\Lcal_{g_1}:Fun(G)
\rightarrow Fun(G)
\en
where $f(g_1g_2)|_{g_2}$ means $f(g_1g_2)$ seen as a function
of $g_2$.
Similarly we can define the right action  on $Fun(G)$ as:
 \eq
(\Rcal_{g_1}f)(g_2)= f(g_2g_1)|_{g_2}
\en
For the basis functions we find easily:
\eq
\Lcal_{g_1} x^{g} = x^{g_1^{-1} g},
~~\Rcal_{g_1} x^{g} = x^{g g_1^{-1}}
\en
Moreover:
 \eqa & &\Lcal_{g_1} \Lcal_{g_2}=\Lcal_{g_1g_2},
~~\Rcal_{g_1} \Rcal_{g_2}=\Rcal_{g_2g_1},\\ & &\Lcal_{g_1}
\Rcal_{g_2}=\Rcal_{g_2} \Lcal_{g_1}
 \ena
  \sk
   \noi {\bf Differential calculus}
    \sk

A {\bf {\sl first-order differential calculus}} on $A$ is defined by
\sk
i) a linear map $d$: $A \rightarrow \Gamma$, satisfying the Leibniz rule
\eq
d(ab)=(da)b+a(db),~~\forall a,b\in A; \label{Leibniz}
\en
The ``space of 1-forms" $\Ga$ is an
appropriate bimodule on $A$, which
essentially means that its elements can be
multiplied on the left and on the right by elements of $A$
[more precisely $A$ is a left module if $\forall a,b \in A, \forall
\rho,\rho' \in \Ga $ we have: $ a(\rho+\rho')=a\rho+a\rho',
~(a+b)\rho=a\rho+b\rho, ~a(b\rho)=(ab)\rho,~ I\rho=\rho$. Similarly
one defines a right module. A left and right module
is a {\sl bimodule} if
$a(\rho b)=(a\rho)b$]. From the Leibniz rule
$da=d(Ia)=(dI)a+Ida$ we deduce $dI=0$.
\sk
ii) the possibility of expressing any $\rho \in \Ga$ as
\eq
\rho=\sum_k a_k db_k \label{adb}
\en
\noi for some $a_k,b_k$ belonging to $A$.
 \sk To build a first
order differential calculus on $Fun(G)$ we need to extend the
algebra $A=Fun(G)$ to a differential algebra of elements
$x^g,dx^g$ (it is sufficient to consider the basis elements and
their differentials). Note however that the $dx^g$ are not
linearly independent. In fact from $0=dI=d(\sumong x^g)=\sumong
dx^g$ we see that only $n-1$ differentials are independent. Every
element $\rho = adb$ of $\Ga$ can be expressed as a linear
combination (with complex coefficients) of terms of the type $x^g
dx^{g'}$. Moreover $\rho b \in \Ga$ (i.e. $\Ga$ is also a right
module) since the Leibniz rule and the multiplication rule
(\ref{mul}) yield the commutations: \eq dx^g x^{g'} = -x^g
dx^{g'}+\de^g_{g'} dx^g
\en
allowing to reorder functions to the left of differentials.
 \sk
 \noi {\bf Partial derivatives}
 \sk
  Consider the differential of a
function $f \in Fun(g)$: \eq df = \sumong f_g dx^g = \sumongnote
f_g dx^g + f_e dx^e= \sumongnote (f_g - f_e)dx^g \equiv
\sumongnote
\part_g f dx^g \label{partcurved}
\en
We have used $dx^e = - \sumongnote dx^g$ (from $\sumong dx^g=0$).
The partial derivatives of $f$ have been defined in analogy with
the usual differential calculus, and are given by \eq
\part_g f = f_g - f_e = f(g) - f(e) \label{partcurved2}
\en
Not unexpectedly, they take here the form of finite differences
(discrete partial derivatives at the origin $e$).
\sk\sk\sk\sk\sk\sk\sk
 \noi
 {\bf Left and right covariance}
\sk A differential calculus is
 {\sl left or right covariant} if the left or right action of
 $G$ ($\Lcal_g$ or $\Rcal_g$) commutes with the exterior derivative $d$.
 Requiring left and right covariance in fact {\sl defines} the action of
 $\Lcal_g$ and $\Rcal_g$ on differentials: $\Lcal_g db \equiv
 d(\Lcal_g b), \forall b \in Fun(G)$ and similarly for
 $\Rcal_g db$. More generally, on elements of $\Ga$
 (one-forms) we define $\Lcal_g$ as:
 \eq
 \Lcal_g (adb) \equiv (\Lcal_g a) \Lcal_g db =
 (\Lcal_g a) d (\Lcal_g b)
 \en
 and similar for $\Rcal_g$.
 Computing for example the left and right action on the differentials
 $dx^g$ yields:
 \eqa
& &\Lcal_g (dx^{g_1})\equiv d(\Lcal_g x^{g_1})=dx^{g^{-1}g_1},~~\\
& &\Rcal_g (dx^{g_1})\equiv d(\Rcal_g x^{g_1})=dx^{g_1 g^{-1}}
\ena
A differential calculus is called {\sl bicovariant} if it is
both left and right covariant.
\sk
\noi
{\bf Left invariant one forms}
 \sk
 As in usual Lie group manifolds, we can introduce a basis in $\Ga$
 of left-invariant one-forms $\theta^g$:
 \eq
  \theta^g \equiv
\sumonh x^{hg} dx^h ~~(=\sumonh x^h dx^{hg^{-1}}),
\label{deftheta}
\en
It is immediate to check that indeed $\Lcal_k \theta^g = \theta^g$. The
relations (\ref{deftheta}) can be inverted:
 \eq
  dx^h = \sumong (x^{hg} - x^h)\theta^g \label{dxastheta}
\en
{} From $0=dI=d\sumong x^g =\sumong dx^g=0$ one finds:
\eq
 \sumong
\theta^g = \sum_{g,h \in G} x^h dx^{hg^{-1}}= \sumonh x^h \sumong
dx^{hg^{-1}}=0 \label{sumtheta}
\en
Therefore we can take as basis of the cotangent space $\Ga$ the
$n-1$ linearly independent left-invariant one-forms $\theta^g$
with $g \not= e$ (but smaller sets of $\theta^g$ can be
consistently chosen as basis, see later).
 \sk
  The $x,~\theta$ commutations (bimodule relations)
are easily derived:
 \eq
  x^h dx^g = x^h \theta^{g^{-1}h} =
\theta^{g^{-1}h} x^g ~~(h\not=g)~~\Rightarrow x^h \theta^g =
\theta^g x^{hg^{-1}}~~(g\not=e) \label{xthetacomm}
\en
implying the general commutation rule between functions and
left-invariant one-forms: \eq f \theta^g = \theta^g \Rcal_g f
\label{fthetacomm}
\en
Thus functions do commute between themselves (i.e. $Fun(G)$ is
a commutative algebra) but do not commute with the
basis of one-forms $\theta^g$. In this sense the differential geometry
of $Fun(G)$ is noncommutative.
\sk
The right
action of $G$ on the elements $\theta^g$ is given by:
\eq
\Rcal_h \theta^g = \theta^{ad(h)g},~~\forall h \in G
\en
where $ad$ is the adjoint action of $G$ on itself, i.e. $ad(h)g
\equiv hgh^{-1}$. Then {\sl bicovariant calculi are in 1-1
correspondence with unions of conjugacy classes (different from
$\{e\}$)} \cite{DMGcalculus}: if $\theta^g$ is set to zero, one must set to
zero all the $\theta^{ad(h)g},~\forall h \in G$ corresponding to the
whole conjugation class of $g$.
 \sk
 We denote by $G'$ the subset corresponding
 to  the union of conjugacy classes
 that characterizes the bicovariant calculus on $G$
 ($G' = \{g \in G |\theta^g \not= 0\}$).
 Unless otherwise indicated, repeated indices are
 summed on $G'$ in the following.
 \sk

A bi-invariant (i.e. left and right invariant) one-form $\Theta$
is obtained by summing on all $\theta^g$ with $g \not= e$: \eq
\Theta = \sumongnote \theta^g
\en
 {\bf Exterior product}
\sk For a bicovariant differential calculus on a Hopf algebra $A$
an {\sl exterior product}, compatible with the left and right
actions of $G$, can be defined by \eq \theta^{g_1} \we
\theta^{g_2}=\theta^{g_1} \otimes \theta^{g_2} - \theta^{g_1^{-1}
g_2 g_1} \otimes \theta^{g_1} \equiv \theta^{g_1} \otimes
\theta^{g_2}- \La^{g_1g_2}_{~~~g_3g_4}~ \theta^{g_3} \otimes
\theta^{g_4}
 \label{exprod}
\en
where the tensor product between elements $\rho,\rhop \in \Ga$
is defined to
have the properties $\rho a\otimes \rhop=\rho \otimes a \rhop$, $a(\rho
\otimes \rhop)=(a\rho) \otimes \rhop$ and $(\rho \otimes \rhop)a=\rho
\otimes (\rhop a)$.

Note that:
 \eq
  \theta^{g} \we \theta^{g}=0~~~~\mbox{(no sum on $g$)}
\en

  Left and right actions on $\Ga \otimes \Ga$ are
  simply defined by:
  \eqa
 & & \Lcal_h (\rho \otimes \rhop)= \Lcal_h \rho \otimes \Lcal_h
  \rhop,~~~\\
& &\Rcal_h (\rho \otimes \rhop)= \Rcal_h \rho \otimes \Rcal_h
  \rhop
  \ena
  (with the obvious generalization to $\Ga \otimes ...\otimes \Ga$)
  so that for example:
\eqa
& &\Lcal_h (\theta^i \otimes \theta^j)=
\theta^i \otimes \theta^j,~~~~\\
& & \Rcal_h (\theta^i \otimes \theta^j)=
\theta^{ad(h)i} \otimes \theta^{ad(h)j}
\ena
 Compatibility  of the exterior product with $\Lcal$ and $\Rcal$
 means that
 \eq
 \Lcal(\theta^i \we \theta^j)=\Lcal\theta^i \we \Lcal
 \theta^j, ~~\Rcal(\theta^i \we \theta^j)=\Rcal\theta^i \we \Rcal
 \theta^j
 \en
 only the second relation being nontrivial, and verifiable using
 the definition (\ref{exprod}).

  We can generalize the definition
(\ref{exprod}) to exterior products of $n$ one-forms:
 \eq
\theta^{i_1} \we ... \we \theta^{i_n} \equiv
\W{i_1..i_n}{j_1..j_n}~
 \theta^{j_1} \otimes ...\otimes
\theta^{j_n}
\en
\noi or in short-hand notation: \eq \theta^{1} \we ... \we
\theta^{n}= W_{1..n}~
 \theta^{1} \otimes ...\otimes
\theta^{n}
\en
\noi where the labels 1..n in $W$ refer to index couples. The
numerical coefficients $W_{1\ldots n}$ are given through the
recursion relation \eq W_{1\ldots n} = {\cal I}_{1\ldots n}
W_{1\ldots n-1} , \label{wedge2}
\en
where \eq {\cal I}_{1\ldots n} = 1 - \Lambda_{n-1,n} +
\Lambda_{n-2,n-1} \Lambda_{n-1,n}   \ldots -(-1)^n \Lambda_{12}
\Lambda_{23} \cdots \Lambda_{n-1,n} \label{wedge3}
\en
$\La$ being defined in (\ref{exprod}) and $W_1 = 1$. The space of
$n$-forms $\Ga^{\we n}$ is therefore defined as in the usual case
but with the new permutation operator $\La$, and can be shown to
be a bicovariant bimodule (see for ex. \cite{Athesis}), with left
and right action defined as for $\Ga \otimes ...\otimes \Ga$ with
the tensor product replaced by the wedge product. \sk \noi {\bf
Exterior derivative} \sk
 Having the exterior product we can define the {\sl exterior
derivative}
\eq
d~:~\Gamma \rightarrow \Gamma \we \Gamma
\en
\eq
d (a_k db_k) = da_k \we db_k,
\en
\noi which can easily be extended to $\Gamma^{\we n}$ ($d:
\Gamma^{\we n} \rightarrow \Gamma^{\we (n+1)}$), and has the
following properties:
\eq
 d(\rho \we \rhop)=d\rho \we \rhop +
(-1)^k \rho \we d\rhop \label{propd1}
\en
\eq
d(d\rho)=0\label{propd2}
\en
\eq
\Lcal_g (d\rho)=d \Lcal_g \rho \label{propd3}
\en
\eq \Rcal_g (d\rho)=d \Rcal_g \rho \label{propd4}
\en
\noi where $\rho \in \Ga^{\we k}$, $\rhop \in \Ga^{\we n}$,
$\Ga^{\we 0} \equiv Fun(G)$ . The last two properties express the
fact that $d$ commutes with the left and right action of $G$. \sk
\noi {\bf Tangent vectors} \sk Using (\ref{dxastheta}) to expand
$df$ on the basis of the left-invariant one-forms $\theta^g$
defines the (left-invariant) tangent vectors $t_g$:
 \eqa
 & & df=\sumong f_g dx^g  = \sumonhp (\Rcal_{h^{-1}} f - f ) \theta^h
\equiv \nonumber \\
& &~~~\equiv \sumonhp (t_h f) \theta^h \label{partflat}
\ena
so that the ``flat" partial derivatives $t_h f$ are given by
\eq
t_h f = \Rcal_{h^{-1}} f - f \label{partflat2}
\en
 The Leibniz rule for the flat partial derivatives $t_g$
reads:
\eq
 t_g (ff')=(t_g f) \Rcal_{g^{-1}} f'  + f t_g f' \label{tgLeibniz}
\en

In analogy with ordinary differential calculus, the operators
$t_g$ appearing in (\ref{partflat}) are called (left-invariant)
{\sl tangent vectors}, and in our case are given by
 \eq
  t_g =
\Rcal_{g^{-1}}- id \label{tangent}
\en
They satisfy the composition rule: \eq t_g t_{g'}= \sum_h
\c{h}{g,g'} t_h \label{chichi}
\en
where the structure constants are:
\eq
\c{h}{g,g'}=\de^h_{g'g} - \de^h_{g}-\de^h_{g'}
\label{cconst}
\en
and are $ad(G)$ invariant: \eq \c{ad(h)g_1}{~~ad(h)g_2,ad(h)g_3}=
\c{g_1}{g_2,g_3} \label{adhc}
\en

{\bf Note 4.1 :}
The exterior derivative on any $f \in Fun(G)$ can be expressed as
a commutator of $f$ with the bi-invariant one-form $\Theta$:
\eq
df = [\Theta , f]
\en
as one proves by using (\ref{fthetacomm}) and (\ref{partflat}).
\sk
 {\bf Note 4.2 :} From the fusion rules (\ref{chichi}) we
deduce the ``deformed Lie algebra" (cf. ref.s \cite{Wor,AC,
Athesis}):
 \eq
  t_{g_1} t_{g_2} -
\L{g_3,g_4}{g_1,g_2}t_{g_3} t_{g_4}= \C{h}{g_1,g_2} t_h
\en
where the $\Cb$ structure constants are given by:
 \eq
\C{g}{g_1,g_2} \equiv \c{g}{g_1,g_2} - \L{g_3,g_4}{g_1,g_2}
\c{g}{g_3,g_4}=\c{g}{g_1,g_2} - \c{g}{g_2,g_2 g_1
g_2^{-1}}=\de^{ad(g_2^{-1})g}_{g_1} - \de^g_{g_1} \label{Cconst}
\en
  and besides property (\ref{adhc}) they also satisfy:
 \eq
\C{g}{g_1,g_2}=\C{g_1}{g,g_2^{-1}} \label{propC}
\en
Moreover the following identities hold:

{\bf i)} {\sl deformed Jacobi identities:}
 \eq  \C{k}{h_1,g_1}
\C{h_2}{k,g_2} - \L{g_3,g_4}{g_1,g_2}\C{k}{h_1,g_3}\C{h_2}{k,g_4}=
\C{k}{g_1,g_2} \C{h_2}{h_1,k} \label{Jacobi} \en

{\bf ii)} {\sl fusion identities:} \eq
\C{k}{h_1,g} \C{h_2}{k,g'}=
\c{h}{g,g'} \C{h_2}{h_1,h} \label{adfusion}
\en

Thus the $\Cb$ structure constants are a representation (the
adjoint representation) of the tangent vectors $t$.
 \sk
  \noi
   {\bf Cartan-Maurer equations, connection and curvature}
 \sk
  From the
definition (\ref{deftheta}) and eq. (\ref{fthetacomm}) we deduce
the Cartan-Maurer equations:
 \eq
 d\theta^g + \sum_{g_1,g_2}
\c{g}{g_1,g_2}\theta^{g_1}\we \theta^{g_2}=0 \label{CM}
\en
where the structure constants $\c{g}{g_1,g_2}$ are those
given in (\ref{cconst}).
\sk
Parallel transport of the vielbein $\theta^g$
can be defined as in
ordinary Lie group manifolds:
\eq
\nabla \theta^g= - \omc{g}{g'} \otimes \theta^{g'}
 \label{parallel}
\en
where $\omc{g_1}{g_2}$ is the connection one-form:
\eq
\omc{g_1}{g_2}= \Gc{g_1}{g_3,g_2} \theta^{g_3}
\en
Thus parallel transport is a map from $\Ga$ to $\Ga \otimes \Ga$;
by definition it must satisfy:
\eq
\nabla (a \rho) = (da)\otimes \rho + a \nabla \rho,~~~\forall a \in
A,~\rho \in \Ga \label{parallel1}
\en
and it is a simple matter to verify that this relation is
satisfied with the usual parallel transport of Riemannian
manifolds. As for the exterior differential, $\nabla$ can be
extended to a map $\nabla : \Ga^{\we n} \otimes \Ga
\longrightarrow \Ga^{\we (n+1)} \otimes \Ga $ by defining:
 \eq
\nabla (\varphi \otimes \rho)=d\varphi \otimes \rho +
 (-1)^n \varphi \nabla
\rho
\en

Requiring parallel transport to commute with the left and right
action of $G$ means:
 \eqa & &\Lcal_{h} (\nabla \theta^{g})=\nabla
( \Lcal_{h} \theta^{g}) =\nabla \theta^g\\ & &\Rcal_{h} (\nabla
\theta^{g})=\nabla ( \Rcal_{h} \theta^{g}) =\nabla \theta^{ad(h)g}
\ena
 Recalling that  $\Lcal_{h} (a \rho)=(\Lcal_h a) (\Lcal_h
\rho)$ and $\Lcal_{h} (\rho \otimes \rho')=(\Lcal_h \rho) \otimes
(\Lcal_h \rho'),~\forall a \in A,~\rho,~\rho' \in \Ga$ (and
similar for $\Rcal_h$),
 and substituting
(\ref{parallel}) yields respectively:
\eq
\Gc{g_1}{g_3,g_2} \in \Cb
\en
and
 \eq
 \Gc{ad(h)g_1}{ad(h)g_3,ad(h)g_2}=\Gc{g_1}{g_3,g_2} \label{adga}
\en
Therefore the same situation arises as in the case of Lie groups,
for which  parallel transport on the group manifold commutes
with left and right action iff the connection components are
$ad(G)$ - conserved constant tensors. As for Lie groups, condition
(\ref{adga}) is satisfied if one takes $\Ga$ proportional to the
structure constants. In our case, we can take any combination of
the $C$ or $\Cb$ structure constants, since both are $ad(G)$
conserved constant tensors. As we see below, the $C$ constants
can be used to define a torsionless connection, while the $\Cb$
constants define a parallelizing connection.

\sk
 As usual, the {\sl curvature} arises from $\nabla^2$:
  \eq
  \nabla^2 \theta^g = - \R{g}{g'} \otimes \theta^{g'}
\en
\eq
\R{g_1}{g_2} \equiv d \omc{g_1}{g_2} + \omc{g_1}{g_3} \we
\omc{g_3}{g_2} \label{curvature}
\en

The {\sl torsion} $R^g$ is defined by:
\eq
R^{g_1} \equiv d\theta^{g_1} +  \omc{g_1}{g_2} \we \theta^{g_2}
\label{torsion}
\en

Using the expression of $\om$ in terms of $\Ga$ and the
Cartan-Maurer equations yields
 \eqa
& & \R{g_1}{g_2} =
 (- \Gc{g_1}{h,g_2}
\c{h}{g_3,g_4} + \Gc{g_1}{g_3,h} \Gc{h}{g_4,g_2})~ \theta^{g_3}
\we \theta^{g_4}=\nonumber \\ & & = (- \Gc{g_1}{h,g_2}
\C{h}{g_3,g_4} + \Gc{g_1}{g_3,h} \Gc{h}{g_4,g_2}- \Gc{g_1}{g_4,h}
\Gc{h}{g_4g_3g_4^{-1},g_2})~\theta^{g_3} \otimes
 \theta^{g_4}\nonumber
\ena \eqa & & R^{g_1}= (- \c{g_1}{g_2,g_3} + \Gc{g_1}{g_2,g_3})~
\theta^{g_2} \we \theta^{g_3}= \nonumber \\ & & (-
\C{g_1}{g_2,g_3} + \Gc{g_1}{g_2,g_3}-
\Gc{g_1}{g_3,g_3g_2g_3^{-1}})\theta^{g_2} \otimes \theta^{g_3}
\ena

Thus a connection satisfying:
 \eq
  \Gc{g_1}{g_2,g_3}-
\Gc{g_1}{g_3,g_3g_2g_3^{-1}}=\C{g_1}{g_2,g_3} \label{rconn}
  \en
corresponds to a vanishing torsion $R^g =0$ and could be
  referred to as a ``Riemannian" connection.
\sk
 On the other hand,  the choice:
   \eq
  \Gc{g_1}{g_2,g_3}=\C{g_1}{g_3,g_2^{-1}} \label{parconn}
  \en
corresponds to a vanishing curvature $\R{g}{g'}=0$, as can be
checked by using the fusion equations (\ref{adfusion}) and
property (\ref{propC}). Then (\ref{parconn}) can be called the
parallelizing connection: {\sl finite groups are parallelizable.}
\sk\sk\sk\sk\sk
 \noi {\bf Tensor transformations } \sk Under the
familiar transformation of the connection 1-form: \eq
(\omc{i}{j})' = \a{i}{k} \omc{k}{l} \ainv{l}{j} + \a{i}{k} d
\ainv{k}{j} \label{omtransf}
\en
the curvature 2-form transforms homogeneously:
\eq
(\R{i}{j})' = \a{i}{k} \R{k}{l} \ainv{l}{j}
\en
The transformation rule (\ref{omtransf}) can be seen as induced by
the change of basis $\theta^i=\a{i}{j} \theta^j$, with $\a{i}{j}$
invertible $x$-dependent matrix (use eq. (\ref{parallel1}) with
$a\rho=\a{i}{j} \theta^j$).
 \sk
  \noi
   {\bf Metric}
\sk
 The metric tensor $\ga$ can be defined as an element of $\Ga
\otimes \Ga$:
 \eq
  \ga = \ga_{i,j} \theta^i \otimes \theta^j
  \en
 Requiring it to be invariant under left and right action of
 $G$ means:
 \eq
 \Lcal_h (\ga)=\ga=\Rcal_h (\ga)
 \en
or equivalently, recalling $\Lcal_h(\theta^i \otimes
\theta^j)=\theta^i \otimes \theta^j$, $\Rcal_h(\theta^i \otimes
\theta^j)=\theta^{ad(h)i}\otimes \theta^{ad(h)j}$  :
 \eq
  \ga_{i,j} \in \Cb,~~  \ga_{ad(h)i,ad(h)j}=\ga_{i,j} \label{gabiinv}
 \en
These properties are  analogous to the ones satisfied by the
Killing metric of Lie groups, which is indeed constant and
invariant under the adjoint action of the Lie group.
 \sk
 On finite $G$ there are various choices of biinvariant
 metrics. One can simply take $\ga_{i,j}=\de_{i,j}$,
   or $\ga_{i,j}= \C{k}{l,i} \C{l}{k,j}$.
   \sk
 For any biinvariant metric $\ga_{ij}$ there are tensor transformations
 $\a{i}{j}$ under which $\ga_{ij}$ is invariant, i.e.:
 \eq
 \a{h}{h'} \ga_{h,k} \a{k}{k'}=\ga_{h',k'} \Leftrightarrow
\a{h}{h'} \ga_{h,k} = \ga_{h',k'} \ainv{k'}{k} \label{ginv}
 \en
 These transformations are simply given by the matrices that
rotate the indices according to the adjoint action of $G$:
 \eq
 \a{h}{h'} (g) = \de^{ad(\al(g))h}_{h'} \label{Gadjoint}
 \en
 where $\al(g): G \mapsto G$ is an arbitrary mapping. Then
 these matrices are functions of $G$ via this mapping, and
 their action leaves $\ga$ invariant because of the its biinvariance
 (\ref{gabiinv}). Indeed
substituting these matrices in (\ref{ginv}) yields:
 \eq
 \a{h}{h'} (g) \ga_{h,k} \a{k}{k'} (g)=
 \ga_{ad([\al(g)]^{-1})h',ad([\al(g)]^{-1})k'}= \ga_{h',k'}
 \en
proving the invariance of $\ga$.

Consider now a contravariant vector $\varphi^i$ transforming as
$(\varphi^i)'=\a{i}{j}(\varphi^j)$. Then using (\ref{ginv}) one
can easily see that
 \eq
  (\varphi^k \ga_{k,i})'=  \varphi^{k'} \ga_{k',i'} \ainv{i'}{i}
  \en
  i.e.  the vector $\varphi_i \equiv \varphi^k \ga_{k,i}$ indeed
  transforms as a covariant vector.
\sk
 \noi
  {\bf Lie derivative and diffeomorphisms}
   \sk
    The notion of diffeomorphisms, or general coordinate transformations,
is fundamental in gravity theories. Is there such a notion in the
setting of differential calculi on finite groups ? The answer is
affirmative, and is based on general results obtained for Hopf algebras
\cite{AC,LCqgravity,Athesis}, of which finite groups are a simple example.
As for differentiable manifolds, it relies on the existence of the
Lie derivative.

Let us review the situation for Lie group manifolds. The Lie
derivative $l_{t_i}$ along a left-invariant tangent vector $t_i$
is related to the infinitesimal right translations generated by
$t_i$:
 \eq
 l_{t_i} \rho = \limepsizero {1\over \epsi} [\Rcal_{\exp [\epsi
 t_i]} \rho - \rho] \label{Liederivative1}
 \en
 $\rho$ being an arbitrary tensor field. Introducing the
 coordinate dependence
 \eq
 l_{t_i} \rho (y) = \limepsizero {1\over \epsi}
  [\rho (y + \epsi t_i) - \rho (y) ]
 \en
identifies the Lie derivative $ l_{t_i}$ as a directional
derivative along $t_i$.  Note the difference in meaning of the
symbol $t_i$ in the r.h.s. of these two equations: a group
generator in the first, and the corresponding tangent vector in
the second.
\sk
 For finite groups the Lie derivative takes the form:
 \eq
 l_{t_g} \rho =  [\Rcal_{g^{-1}} \rho - \rho] \label{Liederivative2}
 \en
so that the Lie derivative along $t_g$ coincides with the tangent vector $t_g$:
 \eq
l_{t_g}=\Rcal_{g^{-1}}-id=t_g
\en
cf. the definition of $t_g$ in (\ref{tangent}). For example
 \eq
l_{t_g} (\theta^{g_1} \otimes \theta^{g_2}) =
 \theta^{ad(g^{-1})g_1} \otimes \theta^{ad(g^{-1})g_2}-
 \theta^{g_1} \otimes \theta^{g_2}
 \en

 As in the case of differentiable manifolds, the Cartan formula
 for the Lie derivative acting on p-forms holds:
\eq
 l_{t_g}= i_{t_g} d + d i_{t_g}
 \en
see ref.s  \cite{AC,Athesis,castGFG,castDCFG}.

 Exploiting this formula, diffeomorphisms
  (Lie derivatives) along generic tangent vectors $V$
 can also be consistently defined via the operator:
\eq
 l_{V}= i_{V} d + d i_{V}
 \en
 This requires
  a suitable definition
 of the contraction operator $i_V$  along generic tangent vectors
 $V$, discussed in ref. \cite{Athesis,castGFG}.

 We have then a way
 of defining ``diffeomorphisms" along arbitrary (and x-dependent)
 tangent vectors for any tensor $\rho$:
 \eq
 \delta \rho = l_V \rho
 \en
and of testing the invariance of candidate lagrangians under the
generalized Lie derivative.
\sk
\noi
{\bf Haar measure and integration}
\sk
Since we want to define actions (integrals on
$p$-forms), we must now define integration of $p$-forms on finite
groups.

 Let us start with integration of functions $f$. We define the integral
  map $h$ as a linear functional $h: Fun(G) \mapsto \Cb$ satisfying the
  left and right invariance conditions:
  \eq
  h(\Lcal_g f)=0=h(\Rcal_g f)
  \en
  Then this map is uniquely determined (up to a normalization constant),
  and is simply given by the ``sum over $G$" rule:
  \eq
  h(f)= \sumong f(g)
  \en

 Next we turn to define the integral of a p-form.
Within the differential calculus we have a basis of left-invariant
1-forms, which may allow the definition of a biinvariant volume
element. In general for a differential calculus with $n$
independent tangent vectors, there is an integer $p  \geq n$ such
that the linear space of $p$-forms is 1-dimensional, and $(p+1)$-
forms vanish identically \footnote{with the exception of $Z_2$,
see Section 4.2}. This means that every product of $p$ basis
one-forms $\theta^{g_1} \we \theta^{g_2} \we ... \we \theta^{g_p}$
is proportional to one of these products, that can be chosen to
define the volume form $vol$:
 \eq
 \theta^{g_1} \we \theta^{g_2} \we ... \we \theta^{g_p}=
 \epsilon^{g_1,g_2,...g_p} vol
 \en
 where $\epsilon^{g_1,g_2,...g_p}$ is the proportionality constant.
 Note that the volume $p$-form is obviously left invariant. We can
  prove that it is also right invariant with the following
  argument. Suppose that $vol$ be given by
  $\theta^{h_1} \we \theta^{h_2} \we ... \we \theta^{h_p}$ where
  $h_1,h_2,...h_p$ are given group element labels. Then the right
  action on $vol$ yields:
  \eq
   \Rcal_g [\theta^{h_1} \we  ... \we
  \theta^{h_p}]=
  \theta^{ad(g)h_1} \we ... \we
  \theta^{ad(g)h_p}=
  \epsilon^{ad(g)h_1,...ad(g)h_p} vol
  \en
  Recall now that the ``epsilon tensor" $\epsilon$ is necessarily
  made out of products of the $\La$ tensor of eq. (\ref{exprod}), defining the wedge
  product. This tensor is invariant under the adjoint action
  $ad(g)$, and so is the $\epsilon$ tensor. Therefore
  $\epsilon^{ad(g)h_1,...ad(g)h_p}=\epsilon^{h_1,...h_p}=1$
  and $\Rcal_g vol = vol$. This can be verified in the $S_3$
  example of \cite{castGFG,castDCFG}, and in the $Z_2$ case
  of next Section.

Having identified the volume $p$-form it is natural to set
 \eq
 \int f vol \equiv h(f) = \sumong f(g) \label{intpform}
 \en
 and  define the integral on a $p$-form $\rho$ as:
 \eqa
 & & \int \rho = \int \rho_{g_1,...g_p}~ \theta^{g_1}
 \we ... \we \theta^{g_p}=  \nonumber \\
 & & \int
\rho_{g_1,...g_p}~\epsilon^{g_1,...g_p} vol \equiv \nonumber \\
& & \equiv ~~~ \sumong
\rho_{g_1,...g_p}(g)~\epsilon^{g_1,...g_p}
  \ena
Due to the biinvariance of the volume form, the integral map $\int
: \Ga^{\we p} \mapsto \Cb$ satisfies the biinvariance conditions:
 \eq
  \int \Lcal_g f = \int f = \int \Rcal_g f
  \en

  Moreover, under the assumption that
  $d(\theta^{g_2} \we ... \we \theta^{g_p})=0$, i.e.
  that any exterior product of $p-1$ left-invariant one-forms $\theta$ is closed,
 the important property holds:
  \eq
  \int df =0
  \en
  with $f$  any $(p-1)$-form: $f=f_{g_2,...g_p}~ \theta^{g_2}
\we ... \we \theta^{g_p}$. This property, which allows
  integration by parts, has a simple proof. Rewrite
  $\int df$ as:
  \eqa
& & \int df= \int (d f_{g_2,...g_p})\theta^{g_2} \we ... \we
\theta^{g_p}+ \nonumber \\
& & + \int  f_{g_2,...g_p} d ( \theta^{g_2} \we ... \we
\theta^{g_p})
  \ena
 The second term in the r.h.s.
 vanishes by assumption. Using now (\ref{partflat}) and
 (\ref{intpform}):
\eqa
 & &\int df=
 \int (t_{g_1} f_{g_2,...g_p})\theta^{g_1}\we \theta^{g_2} \we ...
\we \theta^{g_p} = \nonumber \\
 & &\int
[\Rcal_{g_1^{-1}}f_{g_2,...g_p}-f_{g_2,...g_p}]
\epsilon^{g_1,...g_p} vol = \nonumber \\
 & &\epsilon^{g_1,...g_p}
 \sumong
 [\Rcal_{g_1^{-1}}f_{g_2,...g_p}(g)-f_{g_2,...g_p}(g)]=\nonumber\\
 & & = 0
\ena
 Q.E.D.

\subsection{Bicovariant calculus on $Z_2$}

In this Section we illustrate the general theory on $Z_2$, the
simplest possible example.
 \sk
 \noi {\sl Elements:} $e,u$

 with $u^2=e$.
 \sk
 \noi {\sl Conjugation classes:} $\{e\},\{u\}$.
 \sk
\noi There is therefore only one bicovariant calculus,
corresponding to the only nontrivial congugation class  $\{u\}$,
of dimension 1.
 \sk
 \noi {\sl Basis of functions} on $Z_2$: $\{x^e,x^u\}$. Any function $f$
 can be expanded as: $f=f_e x^e + f_u x^u$.
\sk
 \noi The left action of $Z_2$ on the functions is simply:
 $\Lcal_e = id, \Lcal_u(x^e)=x^u, \Lcal_u(x^u)=x^e$. The right
 action coincides with the left action since $Z_2$ is abelian.
 \sk
 \noi {\sl Partial derivatives:}
 \eq
 df= \sumong f_g dx^g = \sumongnote
f_g dx^g + f_e dx^e=  (f_u - f_e)dx^u \Rightarrow \part_u f =
f_u-f_e
\en

 \noi {\sl Left-invariant one-forms:}
 \eqa
   & & \theta^u=x^u dx^e + x^e dx^u=(x^e-x^u)dx^u,\\
   & & ( \theta^e=x^e dx^e + x^u dx^u=(x^u-x^e)dx^u=-\theta^u)
   \ena
   \noi {\sl Inversion formula:}
   \eq dx^u=(x^e-x^u)\theta^u,~~(dx^e = -dx^u)
   \en
   \noi {\sl Commutations} $x,\theta$:
   \eq f\theta^u = \theta^u \Rcal_u f \Rightarrow x^e \theta^u =
   \theta^u x^u,~~x^u \theta^u = \theta^u x^e \en
   \noi {\sl Commutations} $x,dx$:
   \eq x^e dx^u=dx^u x^u,~~x^u dx^u=dx^u x^e \Rightarrow
   fdx^u=dx^u \Rcal_u f \label{xdxcomm}
   \en
   \noi {\sl Left and right action} on $\theta^u$:
   \eq \Lcal_u \theta^u = \Rcal_u \theta^u = (x^u -
   x^e)dx^e=\theta^u \en
   \noi {\sl Exterior product}:
   \eq \theta^u \we \theta^u=0 \en
   when using the general formula (\ref{exprod}). However the case
   of $Z_2$ is special in this respect: $\theta^u \we
   \theta^u $ can be set different from zero consistently with the
   differential calculus. Indeed for $Z_2$ (and only for this case)
   the whole calculus is consistent
   with the exterior product:
   \eq \theta^u \we \theta^u= \theta^u \otimes \theta^u \en

   For example taking the exterior derivative of both members
   of the commutation relations $x^u dx^u=dx^u x^e$ yields an
   identity (using $d^2=0$ and $dx^e=-dx^u$), and does not imply
   $dx^u \we dx^u =0$. Using then the expression of $dx^u$ in terms
   of $\theta^u$ and the $x,\theta$ commutations, one finds
   \eq
   dx^u \we dx^u = - \theta^u \we \theta^u
   \en
   so that $\theta^u \we \theta^u$ can be different from zero. In
   the case of $Z_N,N>2$ the situation is different since taking
   the exterior derivative of the $x,dx$ commutations
   implies the vanishing of the exterior product of a
   left-invariant one-form with itself: then one has to adopt
   the canonical definition as given in (\ref{exprod}).

   For $Z_2$, we denote the two possibilities calculus I ($dx^u \we dx^u=0$)
   and calculus II ($dx^u \we dx^u \not= 0$).
  \sk
 \noi  {\sl Tangent vector}
  \eq
  t_u f = (\Rcal_u - id)f,~~t_ut_u=(\Rcal_u - id)(\Rcal_u - id)=
  \Rcal_e - 2 \Rcal_u + id=2(id-\Rcal_u)=-2t_u
  \en
 \noi {\sl Cartan-Maurer equations}
 \sk
 \noi Calculus I:
  \eq
 d\theta^u=0
 \en
\noi Calculus II:
  \eq
  d\theta^u= dx^u \we dx^e + dx^e \we dx^u=-2 dx^u \we dx^u= 2 \theta^u \we \theta^u
 \en

 \noi  {\sl Connection}
  \eq
  \omc{u}{u}= \Gc{u}{u,u}\theta^u
  \en
 where $\Gc{u}{u,u}=constant=c$ satisfies left and right
 invariance.
 \sk
 \noi {\sl Curvature and torsion}
 \sk

 \noi Calculus I:
 \eqa
 & & \R{u}{u}=d\omc{u}{u}+\omc{u}{u} \we \omc{u}{u}=c~ d\theta^u
  + c^2~ \theta^u \we \theta^u=0 \\
 & & T^u =d \theta^u + c~\theta^u \we \theta^u=0
 \ena
 Calculus II:
 \eqa
 & & \R{u}{u}=c ~d\theta^u
  + c^2~ \theta^u \we \theta^u = (2c + c^2)~ \theta^u \we \theta^u \\
 & & T^u =d \theta^u + c~\theta^u \we \theta^u=(2+c)~
  \theta^u \we \theta^u
 \ena
 In this case $c=-2$ gives a flat and torsionless connection.
 \sk
\noi  {\sl Integration}
\sk
 For calculus I, the volume form is the one-form $\theta^u$,
 and the integral of a one-form $\rho=\rho_u \theta^u$ is
 simply:
 \eq
 \int \rho = \int \rho_u \theta^u= \int \rho_u vol=
 \sumong \rho_u(g)=\rho_u(e) + \rho_u(u)
 \en
 Integration by parts holds since:
 \eq
 \int df = \int (t_u f)\theta^u=\int [(\Rcal_u - id)f ]vol=
 \sumong (\Rcal_u f - f) (g) =0
 \en
 for $f$ = 0-form.
\sk
 In the special case of $Z_2$, choosing calculus II, there is no
 upper limit to the degree of a $p$-form, since all the products
 $\theta^u \we \theta^u \we ...\theta^u$ are nonvanishing. Then
 any one of these products, being bi-invariant, can be chosen as volume form !
 Supposing to take the $p$-form volume as volume form,
  the integral of a p-form
  $\rho_{u,u,...u} \theta^u \we \theta^u \we ...\theta^u$ is then
   simply $\rho_{u,u,...u}(e)+\rho_{u,u,...u}(u)$.
   Choosing $\theta^u \we \theta^u$ as volume form, we find
 $\int d\sigma \not=0$ (where $\sigma$ is a $1$ form); indeed:
   \eq
   \int d\sigma=\int d(\sigma_u \theta^u) = \int (t_u \sigma_u)
   \theta^u \we \theta^u + 2 \int \sigma_u \theta^u \we \theta^u =
   2 [\sigma_u(e) + \sigma_u(u)]
   \en

   {\bf Note 4.3}:
   Choosing higher volume forms one retrieves the integration by
   parts rule, essentially because an exterior product of two or
   more $\theta^u$'s is closed.


   \subsection{Kaluza-Klein gauge theory on $M_4 \times Z_2$}


In this example we label the $M_4$ coordinates as $x^{\mu}$ and
the $Z_2$ coordinate as $y$. Field theories (and in particular
gauge theories) on discrete spaces have been considered by many
authors. The treatment of this Section is closer in spirit to the works of
\cite{coque,DMgauge,madore,oraf}
\sk
\noi{\bf Calculus on $M_4 \times Z_2$}
\sk

The $y$ coordinate can take the values
$e,u$, and any function $f$ on $M_4 \times Z_2$ is expanded as:
 \eq
 f(x,y)=f_e(x) y^e + f_u(x) y^u
 \en
 where $y^e,y^u$ are defined as usual to be ``dual" to the $Z_2$
 points: $y^e(e)=y^u(u)=1,~y^e(u)=y^u(e)=0$. We will frequently use
 the notation:
 \eq
 \fti\equiv \Rcal_u f = f_u(x) y^e + f_e(x) y^u
 \en
 Thus $\fti$ is obtained from $f$ simply by exchanging its
 components along $y^e,y^u$.

  The only independent $Z_2$ differential
 $dy^u$ will be simply denoted by $dy$.

  Note that
 \eq
 \fti dy=dy~f
 \en
 cf. eq. (\ref{xdxcomm}).

 To define completely the differential geometry on $M_4 \times Z_2$ we
need the rules:
 \eq
 dx^{\mu} \we dy=-dy \we dx^{\mu},~~f  dx^{\mu}= dx^{\mu}~f
\en

 A basis of differentials
 is given by $dx^M=(dx^{\mu},~dy)$, so that any one-form $A(x,y)$
 is expanded as:
 \eq
 A(x,y)= A(x,y)_M dx^M=A_{\mu} (x,y) dx^{\mu} + A_\bu (x,y) dy
 \en
 Finally, integration of a function $f(x,y)$ on $M_4 \times Z_2$ is defined by:
 \eq
 \int_{M_4 \times Z_2} f~vol \equiv \int_{M_4} \sum_{Z_2} f(x,y)~
 d^4x= \int_{M_4}[f_e(x) + f_u(x)] ~ d^4x
 \en

\sk
\noi {\bf Gauge potential}
\sk
Consider now the one-form $A$ to be  the potential 1-form of a
gauge theory: then it must be also matrix valued. For example
in ordinary
Yang-Mills theory, $A(x)=A^I_{\mu} T_I dx^{\mu}$ where $T_I$ are
the generators of the gauge group G in some irreducible
representation.

 As in the usual case, we define G-gauge transformations on
 the potential $A(x,y)$ as:
 \eq
 A'=-(dG)G^{-1} +GAG^{-1} \label{Atransf}
 \en
where $G=G(x,y)$ is a group element is some irrep, depending on the
point $(x,y) \in M_4 \times Z_2$. In components:
 \eq
 A_{\mu}'=-(\part_{\mu}G)G^{-1} +GA_{\mu}G^{-1},~~
 A_{\bu}'=-(\part_{\bu} G)\Gti^{-1} +GA_{\bu}\Gti^{-1}
 \label{Acomptransf}
 \en
the derivative along $y$ being denoted by $\part_{\bu}$. Note that
 \eq
\part_{\bu} f(x,y)=f_u(x) -f_e(x)={1\over 2} (y^e-y^u)(\fti-f) \equiv
J(\fti-f)
 \en
where we have introduced the function $J \equiv {1\over 2} (y^e-y^u)$.
 \sk
 The transformation laws tell us something about the matrix structure
 of the gauge potential $A$. The potential components $A_{\mu}$
 must belong to the Lie
 algebra of G, since $(\part_{\mu}G)G^{-1}\in Lie(G)$. On the other hand $A_{\bu}$
 does not belong to Lie(G) but rather to the {\sl group algebra}
 of G. Indeed $\part_{\bu}G$ is a finite difference of group
 elements, and thus $(\part_{\bu}G)G^{-1}$ belongs to the group
 algebra; then the second eq. in (\ref{Acomptransf}) implies that
 $A_{\bu}$ is matrix valued in the group algebra of G.
 \sk
 For definiteness, we consider unitary groups, so that
 $G^{\dagger}=G^{-1}$. Then $A_{\mu}$ is antihermitian (since the generators
 $T_I$ are antihermitian), while $A_{\bullet}$, being in the $U(N)$ group algebra
 is a sum of $U(N)$ matrices.

  We can consistently incorporate  hermitian conjugation
 in the $M_4 \times Z_2$ - differential calculus  by setting:
 \eqa
 & & (dx^{\mu})^{\dagger}=dx^{\mu},~~(dy)^{\dagger}=dy \\
 & & (f dy )^{\dagger}=dy~ f^{\dagger}
 \ena
 \sk
 \noi {\bf Matter fields}
 \sk

Matter fields $\psi$ are taken to transform in an irrep of G:
 \eq
 \psi ' = G \psi,~~(\psi^{\dagger})'=\psi^{\dagger} G^{\dagger}=
 \psi^{\dagger} G^{-1}
 \en
 and their covariant derivative, defined by
 \eq
 D\psi = d\psi +A\psi,~~D\psi^{\dagger} = d\psi^{\dagger} -\psi^{\dagger} A
 \en
 transforms as it should: $(D\psi)'=G (D\psi),~(D\psi^{\dagger})'=
 (D\psi^{\dagger})G^{-1}$. Requiring compatibility of hermitian
 conjugation with the covariant derivative , i.e.
 $(D\psi)^{\dagger}=D \psi^{\dagger}$, implies:
  \eq
  A^{\dagger}=-A
  \en
  that is, $A$ must be an antihermitian connection. This is
  compatible with its transformation rule (\ref{Atransf}). In
  components the antihermitian condition reads:
  \eq
  A_{\mu}^{\dagger}=-A_{\mu},~~A_{\bu}^{\dagger}=-\Ati_{\bu}
  \label{Acompdagger}
  \en
  \sk
  \noi {\bf Field strength }
  \sk
 The field strength $F$ is formally defined as usual:
 \eq
 F = dA + A \we A
 \en
 so that it transforms as:
 \eq
 F'=G~F~G^{-1}
 \en
 The components of the 2-form $F$ are labelled as follows:
 \eq
 F\equiv F_{MN} ~dx^M \we dx^N \equiv \Fmunu ~dx^{\mu} \we dx^{\nu} +
 2 \Fmubu ~dx^{\mu} \we dy + \Fbubu ~dy \we dy
 \en
 Therefore the $F$ components are given by:
 \eqa
 & & \Fmunu={1\over 2} (\part_{\mu} \Anu - \part_{\nu} \Amu + \Amu \Anu -
 \Anu\Amu) \\
 & & \Fmubu= {1\over 2} (\part_{\mu} \Abu - \part_{\bu} \Amu + \Amu \Abu -
 \Abu\Atimu) \\
 & & \Fbubu= \part_{\bu} \Abu  + \Abu \Atibu
 \ena
 and transform as:
 \eqa
 & & \Fmunu'(x,y)=G(x,y)~ \Fmunu(x,y) ~G^{-1}(x,y) \\
 & & \Fmubu'(x,y)=G(x,y) ~\Fmubu(x,y)~ \Gti^{-1}(x,y) \\
 & & \Fbubu'(x,y)=G(x,y)~ \Fbubu(x,y) ~G^{-1}(x,y) \label{Ftransf}
 \ena
 \sk
 \noi {\bf The gauge action }
 \sk
 Formally the gauge action has the same expression as in the usual
 case:
 \eq
 A_{YM}= \int_{M_4 \times Z_2} Tr_G~ [F_{AB} F^{\dagger}_{AB}] ~vol
 \en
 When expanded into components:
 \eq
  A_{YM}= \int_{M_4 \times Z_2} \sum_{Z_2} Tr_G ~[\Fmunu \Fmunu^{\dagger} +
 2 \Fmubu \Fmubu^{\dagger} + \Fbubu \Fbubu^{\dagger} ] ~d^4x
 \en
 This action is invariant under the G gauge transformations
 (\ref{Ftransf}). We now rewrite it in a suggestive way, by
 introducing the ``link" field $U(x,y)$:
 \eq
 U(x,y) \equiv \unop + J^{-1} \Abu
 \en
 Then
 \eq
 \Fmubu={1\over 2} J ~(\part_{\mu} U + \Amu U - U \Atimu) \equiv {1\over
 2}J~ D_{\mu} U,~~\Fbubu={1\over 4} (\unop -U\Uti) \label{FwithU}
 \en
 Using the transformation rules (\ref{Acomptransf}) one
 finds that the link field $U$ and its covariant derivative vary homogeneously:
 \eq
 U'=G ~U~ \Gti^{-1},~~~~(D_{\mu}U)'=G ~(D_{\mu}U)~ \Gti^{-1}
 \en
 Moreover the antihermiticity of $A$ (\ref{Acompdagger}) implies:
 \eq
 U^{\dagger}=\Uti \label{Udagger}
 \en
 (use $\Jti=-J$). Expanding $U(x,y)$ as $U_e(x) y^e + U_u(x) y^u$,
  relation (\ref{Udagger}) becomes $U_e^{\dagger}=U_u$.
 \sk
 Using the expressions (\ref{FwithU}) for the field strength components finally
 yields the action in the form:
 \eq
 A_{YM}=\int d^4x ~Tr_G~ \sum_{Z_2}~ [\Fmunu \Fmunu + {1\over 16}
 D_{\mu} U (D_{\mu} U)^{\dagger} + {1\over 16} (\unop - UU^{\dagger})^2]
 \en
 The sum on $Z_2$ is easy to perform, and taking into account
 $U_e^{\dagger}=U_u$ we find:
 \eq
 A_{YM}= 2 \int d^4x ~Tr_G~  [\Fmunu \Fmunu + {1\over 16}
 D_{\mu} U (D_{\mu} U)^{\dagger} + {1\over 16} (\unop - UU^{\dagger})^2]
 \label{action}
 \en
 where now $U(x) \equiv U_e(x)$ can be seen as a complex Higgs field, with
 a symmetry-breaking potential. The cyclic property of $Tr_G$ has been used
  to achieve this final form of $A_{YM}$. Moreover we have identified
  for simplicity $\Amu \equiv \Amu(u)=\Amu(e)$ so that the sum on $Z_2$ of
  the usual Yang-Mills term just gives a factor of 2.
  \sk
  \noi {\bf Coupling to fermion matter}
  \sk
  We can add a Dirac term $\Lcal_{Dirac}$ to the integrand of $A_{YM}$:
  \eq
  \Lcal_{Dirac}=Re~[i~\psi^{\dagger} \ga_0 \ga_M D_M \psi]
  \en
  where now the matter field $\psi(x,y)$ is a $d=5$ Dirac spinor
  and has therefore 4 complex spinor components. Splitting the sum
  on the index {\small M}:
  \eq
  \Lcal_{Dirac}=Re~[i~\psi^{\dagger} \ga_0 \ga_\mu D_\mu \psi]+
  Re~[i~\psi^{\dagger} \ga_0 \ga_5 \part_{\bu} \psi
  + i \psi^{\dagger} \ga_0 \ga_5 \Abu \psiti]
  \en
  The first term is just the usual kinetic term in $d=4$; the last
  two terms give:
  \eq
  Re~[-i~J~\psibar \ga_5 \psi + i~J~\psibar \ga_5
  U \psiti~]
  \en
  The first term in square parentheses disappears (since its real part vanishes)
  and the second is:
  \eq
  Re~[i ~\psibar_e \ga_5 U_e \psi_u y^{e}-i~\psibar_u \ga_5 U_u \psi_e
  y^u]
  \en
  Summing on $Z_2$ and redefining $\psi \equiv \psi_e,~\chi \equiv
  i~\ga_5 \psi_u,~~U \equiv U_e$ one finds finally:
  \eq
  \Lcal_{Dirac}=i~( \psibar \ga_{\mu} D_{\mu} \psi +
  \chibar \ga_{\mu} D_{\mu} \chi)
  +  \psibar U \chi + \chibar U^{\dagger} \psi
  \en
  that is a kinetic term for the Dirac fields, and Yukawa
  couplings Higgs-Fermi-Fermi. We emphasize the appearance of
  the correct Higgs couplings to the Fermi fields as an output of the
  Kaluza-Klein mechanism on $M_4 \times Z_2$ rather than an ad hoc
  addition to the Lagrangian. Also, the Higgs sector appears in
  (\ref{action}) with
  the right form of the potential. This provides a nice
  interpretation of the Higgs appearance in the $d=4$ theory in
  terms of a Kaluza-Klein  gauge theory coupled to Dirac fermions
  on $M_4 \times Z_2$. The Higgs field is the component of the
  potential 1-form along the discrete dimension.

  Note that the Kaluza-Klein mechanism on discrete internal spaces
  yields a finite number of fields in $d=4$: there is no infinite tower
  of massive modes ! The ``harmonic" analysis (\ref{fonG}) on finite groups is
  elementary.
 \sk\sk\sk
 {\bf Acknowledgements}
 \sk
 I have benefited from numerous discussions with G. Arcioni, P. Aschieri,
 F. Lizzi and M. Tarlini.


\vfill\eject
\end{document}